\documentclass[fleqn,usenatbib]{mnras}

\usepackage{newtxtext,newtxmath}

\usepackage[T1]{fontenc}

\DeclareRobustCommand{\VAN}[3]{#2}
\let\VANthebibliography\thebibliography
\def\thebibliography{\DeclareRobustCommand{\VAN}[3]{##3}\VANthebibliography}

\usepackage{graphicx}	
\usepackage{amsmath}	
\usepackage{lipsum}  

\newcommand{\clockangle}{$\Phi_{CA}$}
\graphicspath{{./}{figures/}}

\title[]{Switchback Deflections Beyond the Early Parker Solar Probe Encounters}

\author[R. Laker et al.]{
R. Laker$^{1}$\thanks{E-mail: ronan.laker15@imperial.ac.uk},
T. S. Horbury$^{1}$,
L. Matteini$^{1}$,
S. D. Bale$^{2,3}$,
J. E. Stawarz$^{1}$,
L. D. Woodham$^{1}$,
T. Woolley$^{1}$
\\
$^{1}$Imperial College London, Blackett Laboratory, South Kensington, SW7 2AZ\\
$^{2}$Physics Department, University of California, Berkeley, CA 94720-7300, USA\\
$^{3}$Space Sciences Laboratory, University of California, Berkeley, CA 94720-7450, USA\\}

\date{Accepted XXX. Received YYY; in original form ZZZ}

\pubyear{2021}

\begin{document}
\label{firstpage}
\pagerange{\pageref{firstpage}--\pageref{lastpage}}
\maketitle

\begin{abstract}
Switchbacks are Aflv\'enic fluctuations in the solar wind, which exhibit large rotations in the magnetic field direction.
Observations from Parker Solar Probe's (PSP's) first two solar encounters have formed the basis for many of the described switchback properties and generation mechanisms.
However, this early data may not be representative of the typical near-Sun solar wind, biasing our current understanding of these phenomena.
One defining switchback property is the magnetic deflection direction.
During the first solar encounter, this was primarily in the tangential direction for the longest switchbacks, which has since been discussed as evidence, and a testable prediction, of several switchback generation methods.
In this study, we re-examine the deflection direction of switchbacks during the first eight PSP encounters to confirm the existence of a systematic deflection direction.
We first identify switchbacks exceeding a threshold deflection in the magnetic field and confirm a previous finding that they are arc-polarized.
In agreement with earlier results from PSP's first encounter, we find that groups of longer switchbacks tend to deflect in the same direction for several hours.
However, in contrast to earlier studies, we find that there is no unique direction for these deflections, although several solar encounters showed a non-uniform distribution in deflection direction with a slight preference for the tangential direction.
This result suggests a systematic magnetic configuration for switchback generation, which is consistent with interchange reconnection as a source mechanism, although this new evidence does not rule out other mechanisms, such as the expansion of wave modes.

\end{abstract}




\section{Introduction} \label{sec:intro}

Switchbacks are Alfv\'enic fluctuations that represent folds in the magnetic field, rather than changes in magnetic polarity \citep{Balogh1999, Neugebauer2013, Mcmanus2020}, and have been observed sporadically in the fast solar wind with previous spacecraft \citep{Balogh1999,  Horbury2018}.
Following observations from Parker Solar Probe \citep[PSP;][]{Kasper2019,Bale2019}, switchbacks have become an active area of research, with many studies describing their properties during the first two solar encounters \citep{DudokdeWit2020,Horbury2020,Farrell2020, Larosa2021}.

One interesting characteristic identified in this early data was the deflection direction of the magnetic field inside switchbacks.
Although \citet{DudokdeWit2020} demonstrated that the magnetic field deflections were isotropic around the Parker spiral direction \citep{Parker1958}, they did note that the longest switchbacks in the times series data seemed to have a preferential direction.
For an interval during PSP's first solar encounter, \citet{Horbury2020} found that the longest switchbacks deflected in the tangential direction in the same sense as the Sun's rotation.
\citet{Kasper2019} also reported a tangential global flow deflection of the solar wind in the same direction, with the concurrence of these observations leading to the suggestion that they may be related \citep{Fisk2020}.

Interest in switchbacks from the first encounters prompted many theories regarding their creation on the Sun, for example, interchange reconnection \citep{Yamauchi2004,Fisk2020}; coronal jets \citep{Sterling2020}; or flux ropes \citep{Drake2021}. There have also been several mechanisms proposed to create switchbacks in-situ via interactions between solar wind streams \citep{Ruffolo2020,Schwadron2021} or through the growth of various wave modes because of solar wind expansion \citep{Squire2020,Zank2020,Liang2021,Mallet2021,Shoda2021}.
\citet{Schwadron2021} and \citet{Fisk2020} both argue that their super-Parker spiral and interchange reconnection ideas naturally explain both the observed tangential deflections from \citet{Horbury2020} and a global solar wind flow \citep{Kasper2019}.
These systematic deflections are not an explicit prediction of in-situ generation methods, which involve expanding Alfv\'enic waves and turbulence \citep{Squire2020, Mallet2021, Shoda2021}, as suggested by \citet{Fisk2020}.

Therefore, evidence, or lack thereof, for tangential deflections in switchbacks can differentiate between theories of their origin.
Crucially, much of the evidence for this property is based on observations from the first solar encounter, when PSP was connected to a small equatorial coronal hole \citep{Bale2019, Badman2020}.
These conditions are not representative of all observations taken to date.
Therefore, in this paper we revisit the idea of tangential deflections in subsequent solar encounters and aim to remove any ambiguity around this switchback property.

\section{Methodology}\label{sec:method}
\subsection{Definition of a Switchback} \label{sec:method:def}

Following the work of several authors \citep{DudokdeWit2020, Horbury2020, Laker2021, Fargette2021}, we consider switchbacks as deflections of the magnetic field away from the Parker spiral direction .
This represents a stable reference direction that is a based on the physical speed of the solar wind, and is insensitive to large-scale magnetic field deflections associated with the switchbacks themselves.
We estimate the Parker spiral direction using the spacecraft position and the solar wind velocity filtered using a Butterworth filter (of order 2) with a cut-off at $3$ hours. 
In particular, we use proton core fits to the SPAN-i measurements \citep{Woodham2021} for the solar wind velocity estimates, and fill any data gaps using moments of the proton distributions measured by the Solar Probe Cup \citep[SPC,][]{{Kasper2016,Case2020}}.

The smooth velocity profile is not influenced by switchbacks that have much smaller durations than the 3 hour cut-off \citep{DudokdeWit2020}.
A systematic under/over estimation of the Parker spiral direction may introduce unwanted results when investigating any preferential deflection direction of switchbacks.
To avoid this, we compare the Parker spiral direction to periods of manually identified solar wind with minimal switchback activity, finding no systematic offset in either the $\vec{R}$-$\vec{T}$ or $\vec{T}$-$\vec{N}$ planes in the RTN coordinate system\footnote{$\vec{R}$ is radial direction from the Sun to the spacecraft, $\vec{N}$ is the component of the solar north direction perpendicular to $\vec{R}$, and $\vec{T}$ completes the right-handed set}.

In this paper, we define a switchback as a magnetic deflection with an angle greater than 45$^{\circ}$ from the Parker spiral direction, as shown in Fig. \ref{fig:exsb}.
We ignore fluctuations with a duration less than 10 seconds and merge switchbacks that are separated by less than 20 seconds.
In order to include complete observations of the switchbacks, rather than just the measurements at deflections above 45$^{\circ}$, we define the switchback boundaries as the point at which the deflection angle crosses a threshold of 
$30^{\circ}$.
This method may exclude smaller fluctuations from our analysis, but it allows for the identification of individual switchbacks.
In contrast, studying each individual magnetic vector above a certain threshold will provide a distribution heavily weighted towards the longest duration switchbacks.

\subsection{Clock Angle} \label{sec:method:ca}

\begin{figure}
    \includegraphics[width=\columnwidth]{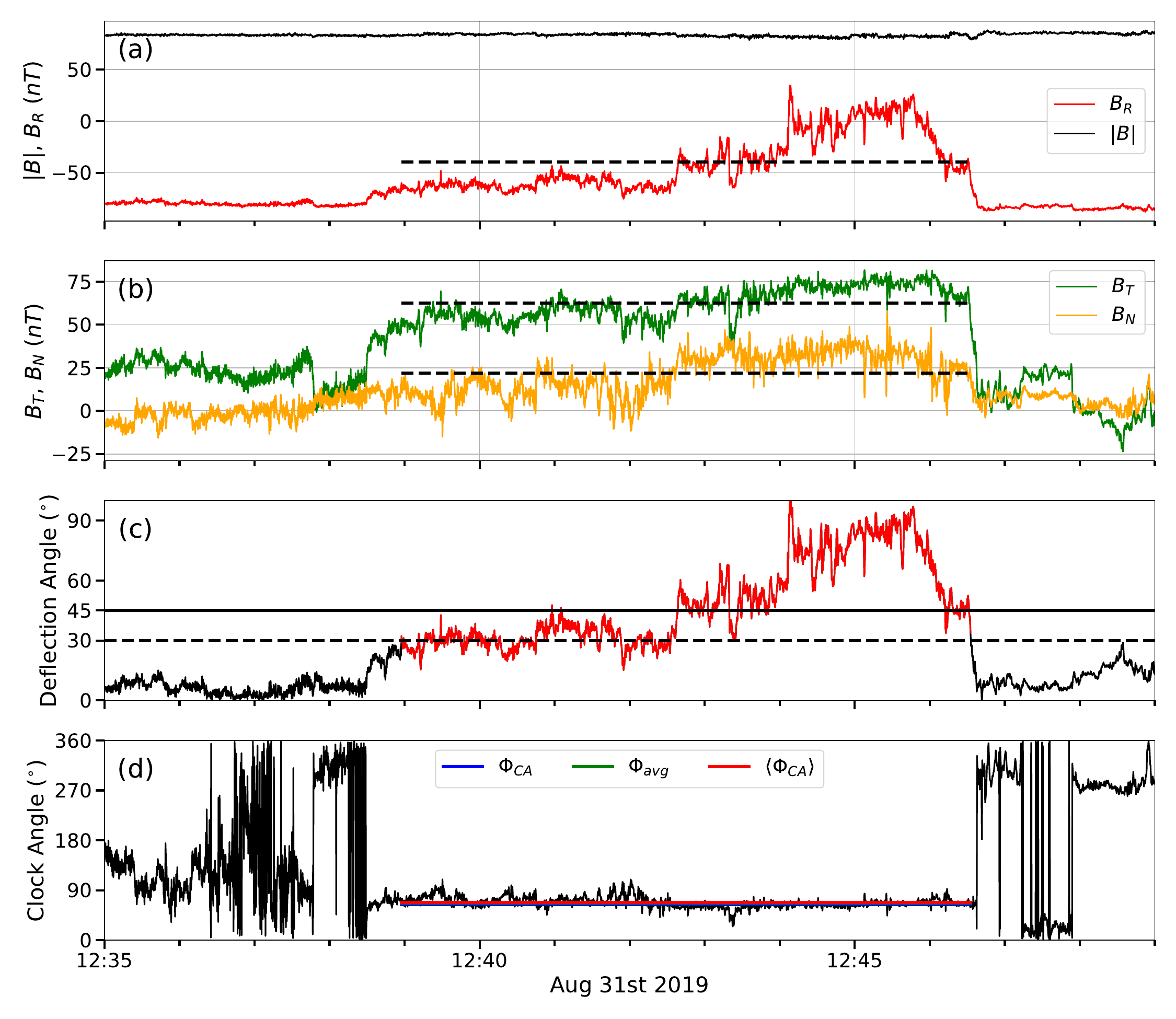}
    \caption{An example switchback identified using the methods outlined in Section \ref{sec:method:def}. (a) and (b) show the magnetic field components, along with a dashed line representing their mean average across the switchback. We show the deflection angle from the Parker spiral in (c), along with our thresholds to define a switchback at $30^{\circ}$ and $45^{\circ}$. (d) shows how the clock angle varies inside a switchback. We then compare different methods for obtaining a single value for clock angle per switchback (defined in the text).}\label{fig:exsb}
\end{figure}

When discussing the deflection direction of switchbacks, we restrict ourselves to magnetic field data from the FIELDS suite \citep{Bale2016}, since rapid deflections in the magnetic field direction can move the proton velocity distribution out of the SWEAP plasma instruments' field of view \citep[e.g., see][]{Woolley2020, Woodham2021}.

Following the work of \citet{Horbury2020} and \citet{Laker2021}, we use the term `clock angle' (\clockangle{}) to describe the direction of switchback deflection.
The clock angle is the angle made by the magnetic field vector in a plane normal to the local Parker spiral direction.
We define $0^{\circ}$, $90^{\circ}$, $180^{\circ}$, $270^{\circ}$ as pointing towards +$\vec{N}$, +$\vec{T}$, -$\vec{N}$, -$\vec{T}$, respectively.
To simplify this procedure, we force the Parker spiral direction to point away from the Sun, which ensures that a deflection with a +$\vec{T}$ component gives $0^{\circ} < \Phi_{CA} < 180^{\circ}$, regardless of the magnetic field polarity.
We can calculate the clock angle for every magnetic field vector in the switchback, meaning there are several methods to measure a single value for \clockangle{} per switchback.
We have opted for a method that averages \clockangle{} when the deflection angle is within $90^{\circ}\pm 10^{\circ}$, or a maximum if this threshold is not reached.
Therefore, we only considered magnetic field vectors towards the centre of the deflection, which avoided averaging issues near the edges, and sampled the magnetic field when it was perpendicular to the local Parker spiral direction, which gave the most consistent results.
Fig. \ref{fig:exsb} demonstrates that there is little difference between our definition of clock angle (\clockangle{}) and the clock angle of the average magnetic components ($\Phi_{avg}$) or the average of the clock angle for each magnetic vector in the switchbacks ($\left\langle \Phi_{CA} \right\rangle$).

\section{Results} \label{sec:results}

\subsection{Arc-Polarized Fluctuations} \label{sec:results:arc}

\begin{figure}
    \includegraphics[width=\columnwidth]{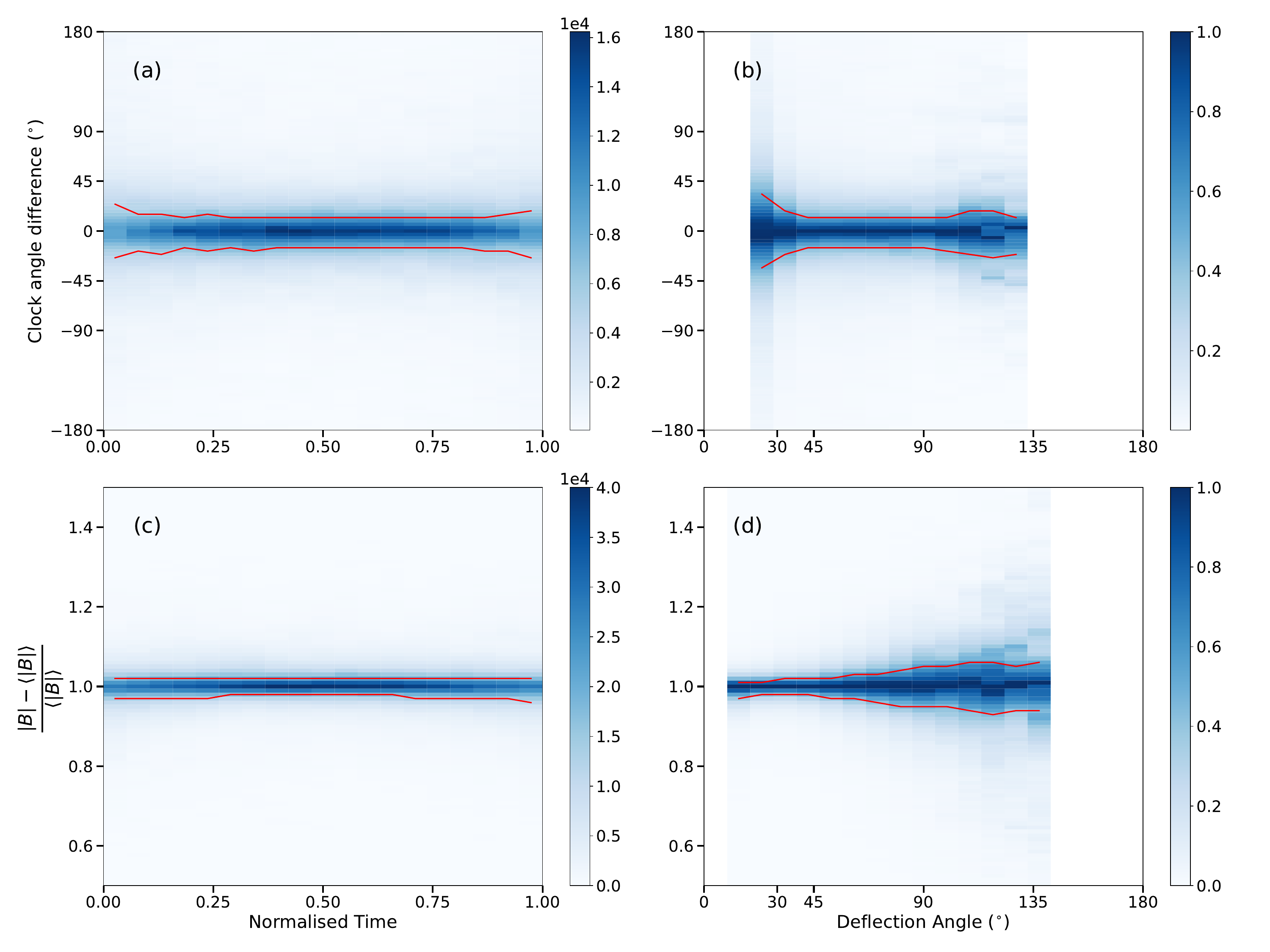}
    \caption{Two dimensional histograms for the magnetic vectors inside all switchbacks over the eight encounters. Top panels show the difference between the 
    clock angle of each vector and our definition of clock angle for an individual switchback (\clockangle{}, from Section \ref{sec:method:ca}) versus normalized time (a) and deflection angle (b). Panels (b) and (d) are normalized by column, which shows that there is little variation in \clockangle{}. Bottom panels show the difference between $|\vec{B}|$ and the mean, which shows that $|\vec{B}|$ is constant inside switchbacks.}\label{fig:arcs}
\end{figure}

Several authors have already stated that switchbacks are `arc-polarized' \citep{Bruno2004,Matteini2014,Horbury2020}, meaning the magnetic field vectors inside a switchback lying on the arc of a sphere of radius $|\vec{B}|$, as demonstrated by the constant \clockangle{} in Fig. \ref{fig:exsb}.
We first test whether this property is consistent across all available encounters, which is a crucial step, since our results are based on the assumption that each switchback has a unique deflection direction.
After identifying switchbacks using the methods in Section \ref{sec:method}, we plot superposed epoch analysis  for $|\vec{B}|$ in the bottom panels of Fig. \ref{fig:arcs}.
Switchbacks show little variation in $|\vec{B}|$ in both normalized time (panel (c)) and deflection angle (panel (d)), which agrees with previous results, confirming the incompressible nature of switchbacks across all encounters \citep{Krasnoselskikh2020, Larosa2021}.

Fig. \ref{fig:arcs} also shows \clockangle{} inside all switchbacks, which again shows little variation compared to our definition.
Therefore, we can conclude that switchbacks are arc-polarized, i.e the fluctuations lie on an arc of the $|\vec{B}|$ sphere and return along the path they deflected from, as shown by the symmetry about $0.5$ normalized time (panel (a)).
Alfv\'enic fluctuations do not have to exhibit this property, and may in principle keep following the arc past $180^{\circ}$ deflection angle, which would case a $180^{\circ}$ change in \clockangle{}.
Such a result may reflect an inherent symmetry of the physical switchback structure, which would be the case for a typical `S-shaped' curve \citep{Kasper2019}.

\subsection{Switchback deflections} \label{sec:results:Consistent}

\begin{figure}
    \includegraphics[width=\columnwidth]{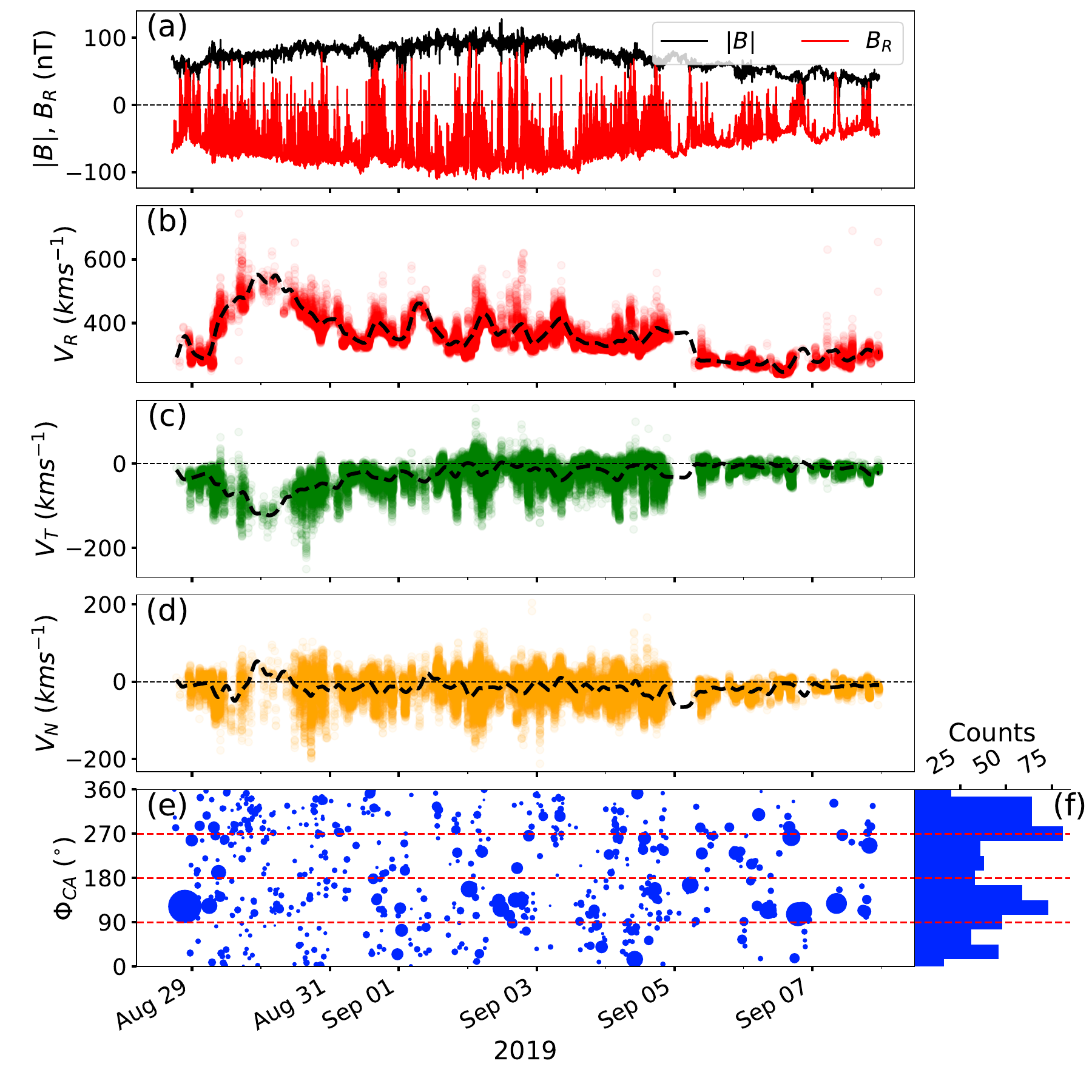}
    \caption{Contextual solar wind plasma data (a-d) for the third PSP encounter, with the \clockangle{} of identified switchbacks plotted in (e), where scatter size is proportional to switchback duration (similar to \citet{Horbury2020}). The histogram in (f) shows the distribution of all switchback deflections in the encounter, which indicates a tendency for deflections around $90^{\circ}$ and $270^{\circ}$ \clockangle{}.}\label{fig:enc}
\end{figure}

Since switchbacks are arc-polarized, with the magnetic field returning along the same path, a single value can describe their deflection direction.
In Fig. \ref{fig:enc}, we plot the deflection direction in terms of \clockangle{} for each switchback in the encounter.
It is clear that an individual switchback can have any value of \clockangle{}, which is a key feature that holds over all the encounters.
However, when considering the distribution of deflection directions across all switchbacks within an encounter, Fig \ref{fig:enc}(f), we find that \clockangle{} is more likely to be around $100^{\circ}$ and $270^{\circ}$.
This anisotropy is also present for the longest switchbacks, which seem to follow the overall distribution.

\begin{figure}
    \includegraphics[width=\columnwidth]{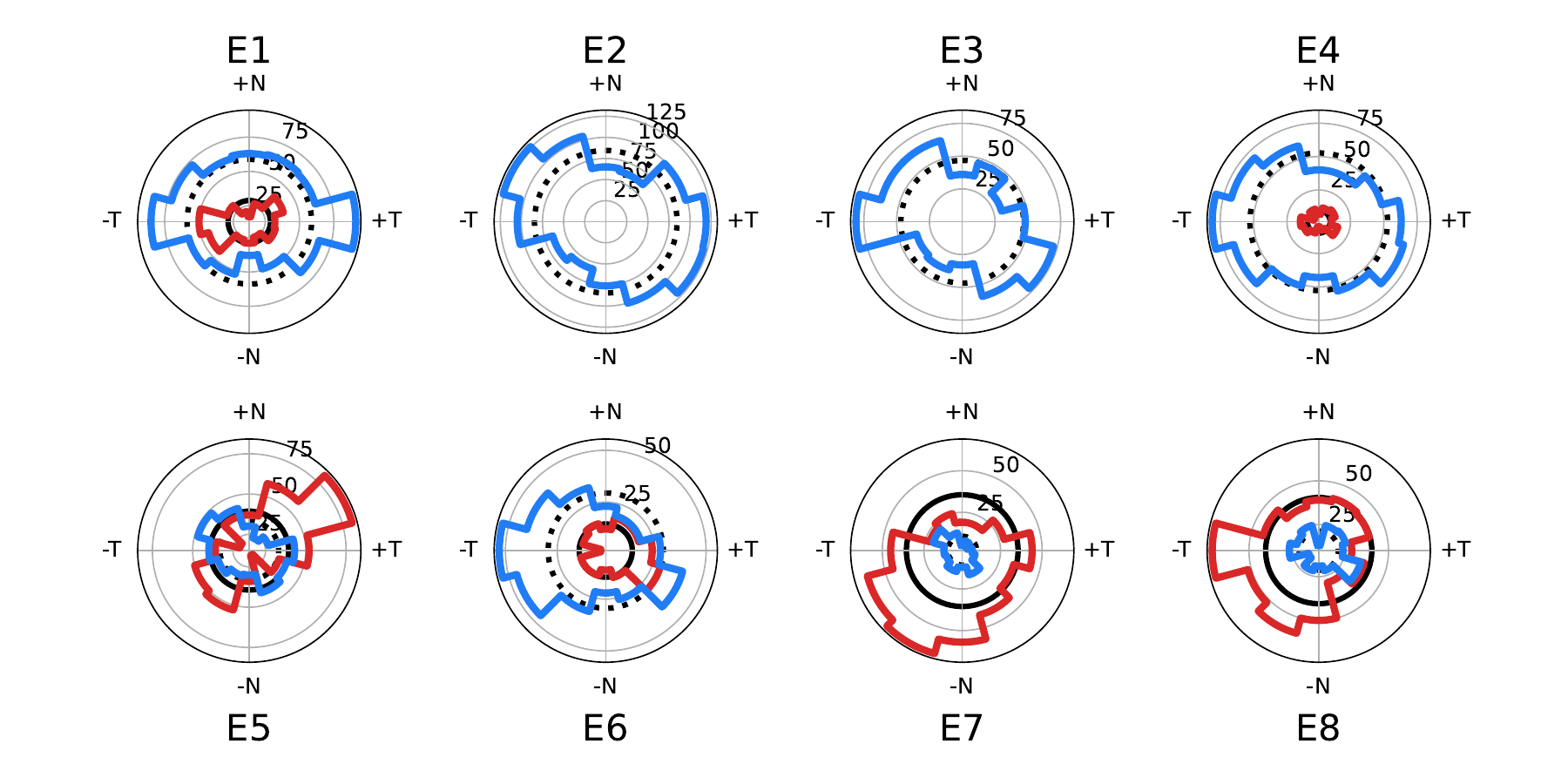}
\caption{Distribution of all switchback clock angles for each PSP encounter, split into positive (red) and negative (blue) polarity groups. We show an expected uniform distribution for both positive (dashed circle) and negative (solid circle) polarities. This demonstrates that switchback deflections are not isotropic, with a preference for deflections in the tangential direction ($90^{\circ}$ and $270^{\circ}$).}\label{fig:polar_hist}
\end{figure}

We summarize the switchback deflections for encounters 1 to 8 by wrapping the histogram in Fig \ref{fig:enc}(f) onto polar axes, as shown in Fig \ref{fig:polar_hist}.
The Alfv\'enic nature of switchbacks allows us to distinguish between positive (red) and negative (blue) polarity background field as this will influence the direction of velocity deflection - i.e. for positive polarity the direction of magnetic deflection is opposite to the velocity deflection, which would rotate the \clockangle{} by $180^{\circ}$.
A null hypothesis of a uniform distribution is plotted for each histogram, which was rejected using Chi-Square test with a 1\% confidence level for all of the observed \clockangle{} distributions across the encounters.
Therefore, we conclude that switchback deflections are not uniformly distributed, with a preference towards tangential, rather than normal, deflections being clear in Fig \ref{fig:polar_hist}.

Interestingly, this preference is not purely tangential (i.e. $90^{\circ}$ or $270^{\circ}$), with negative polarities showing a tendency for an axis of symmetry along $\sim 135^{\circ}$ or $\sim 315^{\circ}$ (e.g. Fig. \ref{fig:polar_hist} encounter 2) and positive having symmetry along $\sim 225^{\circ}$ or $\sim 45^{\circ}$ (e.g. Fig. \ref{fig:polar_hist} encounter 7).
This statement holds true even when using other definitions of clock angle (Sec. \ref{sec:method:ca}).
While the origins of this effect are not clear, it could be related to a systematic magnetic configuration either at the generation site (e.g. a tilted coronal hole boundary), or as the switchback propagates through the solar wind.
It is important to note that not all encounters show this offset symmetry axis (e.g. encounter 1 and 4), suggesting it may be related to the large-scale solar wind structure.

While Fig. \ref{fig:polar_hist} shows the distribution for all switchbacks, Fig. \ref{fig:enc}(e) allows us to see the distribution of the longest switchbacks.

Similar to previous reports \citep{Horbury2020,Laker2021}, these longer switchbacks exhibit `clustering' - i.e. successive longer switchbacks tend to deflect in the same direction.
Such behavior occurs throughout all the encounters, although there is no specific \clockangle{} at which this happens.
During the first encounter, \citet{Horbury2020} showed that longer switchbacks deflected in $+\vec{T}$, which was then subsequently linked to the transverse flow in \citet{Kasper2019}.
However, when looking at the \clockangle{} of switchbacks in encounter 1 (Fig. \ref{fig:polar_hist}) we see that there is no preference for $+\vec{T}$ over $-\vec{T}$, meaning that the earlier result from \citet{Horbury2020} was most likely during a `cluster' of longer switchbacks.

\section{Discussion}\label{sec:discussion}

As \citet{Fisk2020} highlighted, a preference for tangential magnetic deflections is a natural consequence of interchange reconnection \citep{Fisk2001,Fisk2005}.
In this generation mechanism, open magnetic field lines are dragged over closed loops at the Sun, which then reconnect and launch switchbacks into the solar wind.
Therefore, this process is intrinsically linked to the differential rotation of the corona and the solar surface, which would create switchbacks with orientations in the $\pm \vec{T}$ directions \citep[Fig. 1]{Fisk2020}.
Such a mechanism can also explain the clustering behavior in Fig. \ref{fig:enc}, with successive switchbacks being launched from the same reconnection site on the Sun.
A stable magnetic configuration surrounding the reconnection site would then lead to successive switchbacks having similar deflection directions \citep{Zank2020,Fisk2020}, with the distributions in Fig. \ref{fig:polar_hist} reflecting the magnetic field orientation at the Sun.

The super-Parker spiral concept from \citet{Schwadron2021} also predicts a preferential tangential deflection for switchbacks.
Here, the magnetic field foot-point passes from slow to fast solar wind, creating a kink in the magnetic field with a velocity deflection in the same direction as the Sun's rotation. 
Assuming an Alfv\'enic correlation, this would create magnetic field deflections with a $+\vec{T}$ component for negative polarity and $-\vec{T}$ for positive polarity.
Although this was observed during a period of the first solar encounter \citep{Horbury2020}, Fig. \ref{fig:polar_hist} demonstrates that the preference for tangential deflections is not restricted to a single direction (e.g. not only $+\vec{T}$ component for negative polarity).
It is important to note that this argument does not consider the propagation of switchbacks, which could account for a change in orientation.

The observed clustering and preference for tangential deflections are harder to reconcile for those mechanisms that are based on the expansion of randomly orientated in-situ fluctuations \citep{Squire2020,Mallet2021,Shoda2021}.
There would have to be some systematic effect, either in the seeding fluctuations or on the switchbacks as they propagate.
For example, if the Parker spiral could influence the evolution of switchbacks, then this would provide a systematic effect on switchbacks, regardless of their generation mechanism \citep{Johnston2022a,Squire2022a}.

Recently, \citet{Bale2021} and \citet{Fargette2021} demonstrated that patches of switchbacks are organized into structures that are consistent with funnels closer to the Sun, most likely related to super-granules, which can exhibit large-scale magnetic field deflections.
Therefore, it will be interesting to see if simulations can study the influence of large-scale deflections on switchbacks.

However, there are several ways to explain the clustering behavior because all PSP observations are inherently single point measurements.
The most obvious is that there are successive switchbacks with longer duration that have similar magnetic deflection directions.
As discussed in Section \ref{sec:discussion}, this is naturally explained by intermittent events being launched from the Sun, but is harder to explain with expanding waves.
While there are some suggestions that plasma is hotter inside switchbacks \citep{Woodham2021}, there is no agreed plasma signature that can be used to identify switchbacks.
Therefore, with single point measurements, we may misindentify a single coherent structure as several smaller switchbacks in the PSP data.
These `switchbacks' would actually be part of the same larger structure, which could explain the similarity between their deflection directions.
Such an issue could be explored by identifying switchback edges through the presence of wave activity \citep{Agapitov2020,Mozer2020,Larosa2021} and magnetic field dropouts \citep{Farrell2020}, or by finding a unique plasma signature for the inside of a switchback.

The deflection direction of switchbacks is also important when investigating the true size and shape of switchbacks.
\citet{Horbury2020} and \citet{Laker2021} suggested that there is a relationship between the deflection direction of the switchback and the direction in which PSP cuts through it, which arises from the addition of the switchback and spacecraft velocity.
Therefore, the deflection direction can influence how the spacecraft observes the same physical structure, as demonstrated with different cutting angles in \citet{Zank2020}.
If switchbacks have an elongated structure, then the deflection direction will influence the duration of the switchback in the spacecraft time series - i.e., shorter duration seen by PSP when cutting through the side of a long, thin structure \citep{Macneil2020,Laker2021}.
With increasing spacecraft tangential speed in later encounters, it was expected that PSP would travel through the side of switchbacks (assuming they are long and thin \citep{Laker2021}), leading to durations being less affected by deflection direction.
We do not observe this effect in this study, suggesting that the link between deflection direction and duration is not as simple as first thought.
This may be caused by the relationship between an in situ definition of a switchback and the true physical shape of the switchback, and will continue to be a major challenge in the future.

\section{Conclusions}\label{sec:conclusion}

Many of the current switchback generation theories were based on observations from the first few solar encounters from PSP, with the deflection direction of switchbacks being used as evidence.
However, the solar wind conditions in the first encounter were not representative of later encounters, which are now also closer to the Sun.
Therefore, in this study, we have investigated the deflection directions of switchbacks across the first eight encounters.
We have identified switchbacks as deflections in the magnetic field, and confirmed previous results that they have almost constant $|\vec{B}|$ and deflection direction, meaning they can be described as arc-polarized.
We showed that the longer switchbacks still exhibit clustering behavior, where successive switchbacks have similar orientations over periods of hours.
Although switchbacks could deflect in any direction, we demonstrated that they have a tendency to deflect in the tangential, rather than normal, direction.
Such a systematic observation, and the clustering behavior, would be natural consequences of some switchback generation methods, such as interchange reconnection.
More research is needed to rule out other generation theories, such as the evolution of switchbacks and the effect of large-scale magnetic deflections closer to the Sun.
We found no obvious trend in switchback direction or clustering with radial distance from the Sun, although this may be revealed as PSP travels closer in future encounters.

\section*{Acknowledgements}

RL was supported by an Imperial College President's Scholarship, TSH by STFC ST/S000364/1, TW by ST/N504336/1, LDW by ST/S000364/1. JES is supported by the Royal Society University Research Fellowship URF\textbackslash R1\textbackslash 201286. The SWEAP and FIELDS teams acknowledge support from NASA contract NNN06AA01C. This work has made use of the open source and free community-developed space physics packages HelioPy \citep{Stansby2021a} and SpiceyPy \citep{Annex2021}.

\section*{Data Availability}

The data used in this study are available at the NASA Space Physics Data Facility (SPDF): https://spdf.gsfc.nasa.gov.



\bibliographystyle{mnras}
\bibliography{library} 

\bsp	
\label{lastpage}
\end{document}